\documentclass[10pt]{article}
\usepackage[cmex10]{amsmath}
\usepackage{amscd}
\usepackage{amssymb}
\usepackage{amsthm}
\usepackage{relsize}
\usepackage{xspace}

\usepackage{tikz}
\usetikzlibrary{arrows,decorations,backgrounds,shapes}
\newtheorem{thm}{Theorem}[section]
\newtheorem{cor}[thm]{Corollary}
\newtheorem{lem}[thm]{Lemma}

\newtheorem*{lem*}{Lemma}

\theoremstyle{remark}
\newtheorem{remark}[thm]{Remark}
\newtheorem{claim}[thm]{Claim}

\theoremstyle{definition}
\newtheorem{definition}[thm]{Definition}

\theoremstyle{plain}
\newtheorem{fact}[thm]{Fact}

\def\etal{~et~al.\ }

\newcommand{\logic}[1]{\textsf{\upshape\relsize{-0.5}#1}\xspace}
\newcommand{\FP}{\logic{FP}}
\newcommand{\FPC}{\logic{FP+C}}

\newcommand{\FOL}{\logic{FO}}

\newcommand{\ifp}{\operatorname{ifp}}

\newcommand{\spn}{\operatorname{span}}

\newcommand{\cclass}[1]{\ensuremath{\mathrm{#1}}\xspace}
\newcommand{\PTIME}{\cclass{PTIME}}

\newcommand{\LOGSPACE}{\cclass{LOGSPACE}}

\renewcommand {\phi}{\varphi}
\renewcommand{\epsilon}{\varepsilon}
\newcommand{\isom}{\ensuremath{\cong}}

\newcommand{\bigmid}{\;\big|\;}
\newcommand{\modout}{\!\diagup\!\!}
\newcommand{\qedd}[1]{\hfill {\scriptsize\it #1} \ensuremath{\square}}

\def\N{\ensuremath{\mathbb{N}}}

\begin{document}

\title{Capturing Polynomial Time on Interval Graphs}

% \author{\IEEEauthorblockN{\large Bastian Laubner}
% \IEEEauthorblockA{Institut f\"ur Informatik\\
% Humboldt-Universit\"at zu Berlin\\
% laubner@informatik.hu-berlin.de}
% }

\author{Bastian Laubner\\
  {\small Institut f\"ur Informatik}\\
  {\small Humboldt-Universit\"at zu Berlin}\\
  {\small laubner@informatik.hu-berlin.de}
}

\date{%\it\small appeared at the\\ Twenty-Fifth Annual IEEE Symposium on Logic in Computer Science\\ (LICS 2010)
}

\maketitle

\begin{abstract}
  We prove a characterization of all polynomial-time computable queries on the class of interval graphs by sentences of fixed-point logic with counting. More precisely, it is shown that on the class of unordered interval graphs, any query is polynomial-time computable if and only if it is definable in fixed-point logic with counting. This result is one of the first establishing the capturing of polynomial time on a graph class which is defined by forbidden induced subgraphs. For this, we define a canonical form of interval graphs using a type of modular decomposition, which is different from the method of tree decomposition that is used in most known capturing results for other graph classes, specifically those defined by forbidden minors. The method might also be of independent interest for its conceptual simplicity. Furthermore, it is shown that fixed-point logic with counting is not expressive enough to capture polynomial time on the classes of chordal graphs or incomparability graphs.
\end{abstract}

% \begin{keywords}
% interval graphs; capturing of polynomial time; fixed-point logic with counting; canonical forms; modular decomposition
% 
% \end{keywords}

%% start the paper here:
\section{Introduction}\label{sec:intro}

Capturing results in descriptive complexity match the expressive power of a logic with the computational power of a complexity class. The most important open question in this area is whether there exists a natural logic whose formulas precisely define those queries which are computable in polynomial time (\PTIME). While Immerman and Vardi showed in 1982 that fixed-point logic captures \PTIME under the assumption that a linear order is present in each structure (cf. Theorem \ref{thm:immermanVardi}), there is no logic which is currently believed to capture \PTIME on arbitrary unordered structures. Despite that limitation, precise capturing results for \PTIME in the unordered case can be obtained for restricted classes of structures. Since all relational structures of a fixed finite vocabulary can be encoded efficiently as simple graphs, capturing results on restricted graph classes are of particular interest in this context.

This approach has been very fruitful in the realm of graph classes defined by lists of forbidden minors. Most of these results show that \PTIME is captured by fixed-point logic with counting \FPC when restricting ourselves to one such class, such as planar graphs~\cite{gro98a}, graphs of bounded tree-width~\cite{gromar99}, or $K_5$-free graphs~\cite{grohe08definable}. In fact, Grohe has recently shown that \FPC captures \PTIME on any graph class which is defined by a list of forbidden minors~\cite{grohe10fixed}.

Given such deep results for classes of minor-free graphs, it is natural to ask if similar results can be obtained for graph classes which are defined by a (finite or infinite) list of forbidden induced subgraphs. Much less is known here. For starters, it is shown in \cite{grohe09fixed-point} and in Section \ref{sec:noCaptureCompGraphs} that a general capturing result analogous to Grohe's is not possible for \FPC on subgraph-free graph classes, such as chordal graphs or graphs whose complements are comparability graphs of partial orders. These two superclasses of interval graphs are shown to be a ceiling on the structural richness of graph classes on which capturing \PTIME requires less effort than for general graphs. % It is shown here that capturing \PTIME on the class of comparability graphs is essentially just as hard as capturing \PTIME on the class of all graphs.

\begin{thm}\label{thm:notCaptureCompGraphs}
 \FPC fails to capture \PTIME on the class of incomparability graphs and on the class of chordal graphs.
\end{thm}

% One approach towards finding such a match is to consider restricted classes of structures on which such a correspondence can more easily be established. Whereas there has been a lot of research and deep results for classes of graphs defined by excluded minors, little is known for graph classes defined by forbidden induced subgraphs.

The main result in this paper is a positive one affirming that \FPC captures \PTIME on the class of interval graphs. This means that a subset $\mathcal K$ of the class of interval graphs is decidable in \PTIME if and only if there is a sentence of \FPC defining $\mathcal K$.

\begin{thm}\label{thm:captureIntGraphs}
 \FPC captures \PTIME on the class of interval graphs.
\end{thm}

The result is shown by describing an \FPC-definable canonization procedure for interval graphs, which for any interval graph constructs an isomorphic copy on an ordered domain. The capturing result then follows from the Immerman-Vardi theorem. The proof of Theorem \ref{thm:captureIntGraphs} also has a useful corollary.

\begin{cor}\label{cor:IntGraphsDefinable}
 The class of interval graphs is \FPC-definable.
\end{cor}

There has been persistent interest in the algorithmic aspects of interval graphs in the past decades, also spurred by their applicability to DNA sequencing (cf. \cite{zhang94algorithm}) and scheduling problems (cf. \cite{moehring84algorithmic}). In 1976, Booth and Lueker presented the first recognition algorithm for interval graphs~\cite{booth76testing} running in time linear in the number of vertices and edges, which they followed up by a linear-time interval graph isomorphism algorithm~\cite{lueker79linear}. These algorithms are based on a special data structure called \emph{PQ-trees}. Using so-called perfect elimination orderings, Hsu and Ma~\cite{hsu99fast} and Habib\etal~\cite{habib00lex-bfs} later presented linear-time recognition algorithms based on simpler data structures.

All these approaches have in common that they make inherent use of an underlying order of the graph, which is always available in \PTIME computations as the order in which the vertices are encoded on the worktape. Particularly, the construction of a perfect elimination ordering by lexicographic breadth-first search needs to examine the children of a vertex in some fixed order. However, such an ordering is not available when defining properties of the bare unordered graph structure by means of logic. Therefore, most of the ideas developed in these publications cannot be applied in the canonization of interval graphs in \FPC.

We note that an algorithmic implementation of our method would be inferior to the existing linear-time algorithms for interval graphs. Given that our method must rely entirely on the inherent structure of interval graphs and not on an additional ordering of the vertices, we reckon that is the price to pay for the disorder of the graph structure.

The main commonality of existing interval graph algorithms and the canonical form developed here is the construction of a \emph{modular decomposition} of the graph. Modules are subgraphs which interact with the rest of the graph in a uniform way, and they play an important algorithmic role in the construction of modular decomposition trees (cf. \cite{brandstaedt99graph}). As a by-product of the approach in this paper, we obtain a specific modular decomposition tree that is \FPC-definable. Such modular decompositions are fundamentally different from \emph{tree decompositions}, which are the ubiquitous tool of \FPC-canonization proofs for the aforementioned minor-free graph classes (cf. \cite{grohe08definable} for a survey of tree decompositions in this context). Since tree decompositions do not appear to be very useful for defining canonical forms on subgraph-free graph classes, showing the definability of modular decompositions is a contribution to the systematic study of capturing results on these graph classes.

% In 1985, Kozen\etal~\cite{kozen85NC} presented an \cclass{NC} parallel algorithm for computing an interval representation of an interval graph. Generally, parallel algorithms are much more apt for adaptation into \FPC-definable procedures, since their parallel execution on parts of the input requires them to behave in an order-invariant manner to a certain degree. However, their approach does not translate into a graph isomorphism algorithm, and the fact that they

% The methods employed for this canonization easily extend to the larger class of circular-arc graphs.
%
% \begin{thm}\label{thm:captureCircArcGraphs}
%  \FPC captures \PTIME on the class of circular-arc graphs.
% \end{thm}

\section{Preliminaries and notation}

We write $\N$ and $\N_0$ for the positive and non-negative integers, respectively. For $m,n \in \N_0$, let $[m,n] := \{\ell \in \N_0 \bigmid m \leq \ell \leq n \}$ be the \emph{closed interval of integers from $m$ to $n$}, and let $[n] := [1,n]$. Tuples of variables $(v_1, \ldots, v_k)$ are often denoted by $\vec v$ and their length by $|\vec v|$.

%\subsection{Orders}

A binary relation $<$ on a set $X$ is a \emph{strict partial order} if it is irreflexive and transitive. %, i.e., there is no $x \in X$ with $x < x$ and whenever $x < y$ and $y < z$, then also $x < z$. Irreflexivity and transitivity together imply antisymmetry, i.e., whenever $x < y$, it does not hold that $y < x$.
Two elements $x,y$ of a partially ordered set $X$ are called \emph{incomparable} if neither $x < y$ nor $y < x$. We call $<$ a \emph{strict weak order} if it is a strict partial order, and in addition, incomparability is an equivalence relation, i.e., whenever $x$ is incomparable to $y$ and $y$ is incomparable to $z$, then $x$ and $z$ are also incomparable. If $x,y$ are incomparable with respect to a strict weak order $<$, then $x < z$ implies $y < z$.

Finally, a \emph{(strict) linear order} is a strict partial order in which no two elements are incomparable. If $<$ defines a strict weak order on $X$ and $\sim$ is the equivalence relation defined by incomparability, then $<$ induces a linear order on $X \modout \sim$.

\subsection{Graphs}

All graphs in this paper are assumed to be finite, simple, and undirected unless explicitly stated otherwise. Let $G = (V,E)$ be a graph with vertex set $V$ and edge set $E$. Generally, $E$ is viewed as a binary relation. Sometimes, we also find it convenient to view edges $e$ as sets containing their two endpoints, as in $e = \{u,v\} \subseteq V$. For isomorphic graphs $G$ and $H$ we write $G \isom H$.

For $W \subseteq V$ a set of vertices, $G[W]$ denotes the \emph{induced subgraph} of $G$ on $W$. The \emph{neighborhood} of a vertex $v \in V$, denoted $N(v)$, is the set of vertices adjacent to $v$ under $E$, \emph{including $v$ itself}.

A subset $W\subseteq V$ is called a \emph{clique} of $G$ if $G[W]$ is a complete graph. A clique $W$ of $G$ is \emph{maximal} if it is inclusion-maximal as a clique in $G$, i.e., if no vertex $v\in V \setminus W$ can be added to $W$ so that $W\cup \{v\}$ forms a clique. Since maximal cliques are central to the constructions in this paper, they will often just be called \emph{max cliques}. \emph{Cycles} in a graph are defined in the usual way.

A graph $G$ is a \emph{split graph} if its vertex set can be partitioned into two sets $U$ and $V$ so that $G[U]$ is a clique and $V$ is an independent set. We write $G = (U\dot\cup V, E)$ to emphasize on the partition. Similarly, if $G$ is a \emph{bipartite graph}, then we write $G = (U\dot\cup V, E)$ in order to emphasize that $U$ and $V$ are independent sets.

%Similarly, if $G$ is a \emph{split graph}, writing $G = (U\dot\cup V, E)$ emphasizes that $G[U]$ is a clique and $V$ is an independent set.

The main result of this paper, Theorem \ref{thm:captureIntGraphs}, is about interval graphs, which we define and discuss now. The properties of interval graphs mentioned here are based on \cite{gilmore64characterization}.

\begin{definition}[Interval graph]
Let $\mathcal I$ be a finite collection of closed intervals $I_i = [a_i,b_i] \subset \N$. The graph $G_{\mathcal I} = (V,E)$ defined by $\mathcal I$ has vertex set $V = \mathcal I$ and edge relation $I_iI_j \in E :\Leftrightarrow I_i \cap I_j \neq \emptyset$. $\mathcal I$ is called an \emph{interval representation} of a graph $G$ if $G \isom G_{\mathcal I}$. A graph $G$ is an \emph{interval graph} if there is a collection of closed intervals $\mathcal I$ which is an interval representation of $G$.
\end{definition}

If $v \in V$, then $I_v$ denotes the interval corresponding to vertex $v$ in $\mathcal I$. An interval representation $\mathcal I$ for an interval graph $G$ is called \emph{minimal} if the set $\bigcup \mathcal I \subset \N$ is of minimum size over all interval representations of $G$. Any interval representation $\mathcal I$ can be converted into a minimal interval representation by removing a subset of $\N$ from all intervals in $\mathcal I$ (and then considering the remaining points in $\bigcup \mathcal I$ as an initial segment of $\N$).

%embedding the resulting collection of sets consecutively back into $\N$ in an order-preserving manner.

%using the map $n \in \bigcup \mathcal I \setminus K \mapsto | \{ m \in \bigcup \mathcal I \setminus K \bigmid m \leq n \} |$.

If $\mathcal I = \{I_i\}_{i \in [n]}$ is a minimal interval representation of $G$, then there is an intimate connection between the max cliques of $G$ and the sets $M(k) = \{ I_i \bigmid k \in I_i \}$ for $k\in \N$. In fact, if $M(k) \neq \emptyset$ for some $k$, then $M(k)$ forms a clique which is maximal by the minimality condition on $\mathcal I$. Conversely, if $M$ is a max clique of $G$, then $\bigcap_{v\in M} I_v = \{k\}$ for some $k\in\N$ by the minimality of $\mathcal I$, and $M(k) = M$. Thus, a connected graph $G$ is an interval graph if and only if its max cliques can be arranged as a path so that each vertex of $G$ is contained in consecutive max cliques. In this way, any minimal interval representation $\mathcal I$ of $G$ induces an ordering $\lhd_{\mathcal I}$ of $G$'s max cliques. We call a max clique $C$ a \emph{possible end} of $G$ if there is a minimal interval representation $\mathcal I$ of $G$ so that $C$ is $\lhd_{\mathcal I}$-minimal.

Interval graphs are a classical example of an \emph{intersection graph class} of certain objects. Intersection graphs have as vertices a collection $\{o_1,\ldots, o_k\}$ of these objects with an edge between $o_i$ and $o_j$ if and only if $o_i \cap o_j \neq \emptyset$. Notice that any finite graph is the intersection graph of some collection of sets from $\N$, which is not the case when we restrict the allowed sets to intervals.

If $\mathcal Y$ is an intersection graph class, $G =(V,E) \in \mathcal Y$, and $U$ is any subset of $V$, then $G[U]$ is also a member of $\mathcal Y$ since it is just the intersection graph of the objects in $U$. Any graph class $\mathcal G$ that is closed under taking induced subgraphs can also be defined by a possibly infinite list of \emph{forbidden induced subgraphs}, by taking all those graphs not in $\mathcal G$ that are minimal with respect to the relation of being an induced subgraph. A complete infinite family of forbidden induced subgraphs defining the class of interval graphs is given by Lekkerkerker and Boland in \cite{lekkerkerker62representation}.

%The unique number $k\in\N$ corresponding to a maximal clique $M$ in $G$ with respect to the minimal interval representation $\mathcal I$ is also denoted by $M^{\mathcal I}$. %Maximal cliques, or \emph{max cliques} for short, play an important role in the proof of Theorem \ref{thm:captureIntGraphs}.

Some further classes of graphs are important for this paper, and will be defined now.

\begin{definition}[Chordal graph]
A graph is called \emph{chordal} if all its induced cycles are of length 3.
\end{definition}

It is easy to show that every interval graph is chordal. Chordal graphs can alternatively be characterized by the property that the maximal cliques can be arranged in a forest $T$, so that for every vertex of the graph the set of max cliques containing it is connected in $T$ (cf. \cite{diestel06graphtheory}).

\begin{definition}[Comparability graph]
A graph $G = (V,E)$ is called a \emph{comparability graph} if there exists a strict partial ordering $<$ of its vertex set $V$ so that $uv \in E$ if and only if $u,v$ are comparable with respect to $<$.
% Let $G = (V,E)$ be a graph. An \emph{orientation} of the edges of $G$ is a set $O \subset V^2$ so that whenever $(u,v) \in O$, then $uv \in E$, and whenever $uv \in E$, then either $(u,v) \in O$ or  $(v,u) \in O$, but not both. Such an orientation $O$ of $G$ is called \emph{transitive} if whenever $(u,v), (v,w) \in O$, then also $(u,w) \in O$. $G$ is called a \emph{comparability graph} if there exists a transitive orientation of its edges.
\end{definition}

A graph is called an \emph{incomparability graph} if its complement is a comparability graph. It is a well-known fact that every interval graph is an incomparability graph. In fact, a graph is an interval graph if and only if it is a chordal incomparability graph \cite{gilmore64characterization, golumbic04algorithmic}.

% \begin{definition}[Circular-arc graph]
% Let $\mathcal I$ be a finite collection of closed intervals $I_i = [a_i,b_i] \subset \N$ and let $m\in \N$. The graph $G_{\mathcal I\modout m} = (V,E)$ defined by $(\mathcal I, m)$ has vertex set $V = \mathcal I$ and edge relation $I_iI_j \in E :\Leftrightarrow$ there are $n\in I_i$ and $n'\in I_j$ so that $n = n' \mbox{ mod } m$. $\mathcal I$ is called a \emph{circular-arc representation} of a graph $G$ if $G \isom G_{\mathcal I\modout m}$. A graph $G$ is a \emph{circular-arc graph} if there is a collection of closed intervals $\mathcal I$ and an integer $m \in \N$ so that $(\mathcal I, m)$ is a circular-arc representation of $G$.
% \end{definition}
%
% Clearly, any interval graph $G$ is a circular-arc graph, since if $\mathcal I$ is an interval representation of $G$, then $(\mathcal I, \max \bigcup \mathcal I)$ is a circular-arc representation of $G$. However, cycles of any length are circular-arc graphs, but not interval graphs since they are not chordal.

\subsection{Logics}

We assume basic knowledge in logic, particularly of first-order logic \FOL. All structures considered in this paper are graphs $G = (V,E)$, i.e., relational structures with universe $V$ and one binary relation $E$ which is assumed to be symmetric and irreflexive. This section will introduce the fixed-point logics \FP and \FPC. Detailed discussions of these logics can be found in \cite{ebbinghaus99finite, graedel07finite, immerman99descriptive}.

If $\phi$ is a formula of some logic, we write $\phi(x_1, \ldots, x_k)$ to indicate that the free variables of $\phi$ are among $x_1, \ldots, x_k$. If $v_1, \ldots, v_k$ are vertices of a graph $G$, then $G \models \phi[v_1, \ldots, v_k]$ denotes that $G$ satisfies $\phi$ if $x_i$ is interpreted as $v_i$ for all $i \in [k]$. Furthermore, $\phi^G[v_1, \ldots, v_{k-1},\cdot]$ denotes the subset of vertices $v_k$ in $G$ for which $G \models \phi[v_1, \ldots, v_k]$, and similarly, $\phi^G[\cdot,\ldots,\cdot] = \{ \vec v \in V^k \bigmid G \models \phi[\vec v] \}$.

\emph{Inflationary fixed-point logic} \FP is the extension of \FOL by a fixed-point operator with inflationary semantics, which is defined as follows. Let $G=(V,E)$ be a graph, let $X$ be a \emph{relation variable} of arity $r$, and let $\vec x$ be a vector of $r$ variables. Let $\phi$ be a formula whose free variables may include $X$ as a free relation variable and $\vec x$ as free (vertex) variables. For any set $F \subseteq V^r$, let $\phi[F]$ denote the set of $r$-tuples $\vec v \in V^r$ for which $\phi$ holds when $X$ is interpreted as $F$ and $\vec v$ is assigned to $\vec x$. Let the sets $F_i$ be defined inductively by $F_0 = \phi[\emptyset]$ and $F_{i+1} = F_i \cup \phi[F_i]$.  Since $F_i \subseteq F_{i+1}$ for all $i\in \mathbb N_0$, we have $F_k = F_{|V|^r}$ for all $k \geq |V|^r$. We call the $r$-ary relation $F_{|V|^r}$ the \emph{inflationary fixed-point} of $\phi$ and denote it by $\left( \ifp_{X \leftarrow \vec x} \phi \right)$. \FP denotes the extension of \FOL with the $\ifp$-operator.

In 1982, Immerman~\cite{immerman82bounds} and Vardi~\cite{vardi82complexity} showed that \FP characterizes \PTIME on classes of ordered structures\footnote{In fact, Immerman and Vardi showed this capturing result using a different fixed-point operator for \emph{least fixed points}. Inflationary and least fixed points were shown to be equivalent by Gurevich and Shelah~\cite{gurevich85fixed-point} and Kreutzer~\cite{kreutzer04equivalence}. Also, Immerman and Vardi proved the result for general relational structures with an ordering, while we only state their theorem for graphs.}.

\begin{thm}[Immerman-Vardi]\label{thm:immermanVardi}
Let $\mathcal K$ be a class of ordered graphs, i.e., graphs with an additional binary relation $<$ which satisfies the axioms of a linear order. Then $\mathcal K$ is \PTIME-decidable if and only if there is a sentence of \FP defining $\mathcal K$.
\end{thm}

When no ordering is present, then \FP is not expressive enough to capture \PTIME; in fact, it cannot even decide the parity of the underlying vertex set's size. For the capturing result in this paper, we will also need a stronger logic which is capable of such basic counting operations.

For this, let $G = (V,E)$ be a graph and let $N_V := [0,|V|] \subset \N_0$. Instead of $G$ alone, we consider the two-sorted structure $G^+ := (V, N_V, E, <)$ with universe $V \dot\cup N_V$, so that $E$ defines $G$ on $V$ and $<$ is the natural linear ordering of $N_V \subset \N_0$. Notice that $E$ is not defined for any numbers from $N_V$, and also, $<$ does not give any order on $V$. Now we define \FP-sentences on $G^+$ with the convention that all variables are implicitly typed. Thus, any variable $x$ is either a vertex variable which ranges over $V$ or a numeric variable which ranges over $N_V$.

The connection between the vertex and the number sort is established by \emph{counting terms} of the form $\# x \, \phi$ where $x$ is a vertex variable and $\phi$ is a formula. $\# x \, \phi$ denotes the number from $N_V$ of vertices $v \in V$ so that $G \models \phi[v]$. \FPC is now obtained by extending \FP in the two-sorted framework with counting terms. % Notice that each graph $G$ is extended to $G^+$ in a canonical fashion, therefore we can view \FPC-sentences as defining properties of plain graphs.

We can encode numbers from $[0,|N_V|^k - 1]$ with $k$-tuples of number variables. With the help of the fixed-point operator, we can do some meaningful arithmetic on these tuples, such as addition, multiplication, and counting the number of tuples $\vec x$ satisfying a formula $\phi(\vec x)$ (cf. \cite{graedel07finite}).

With its power to handle basic arithmetic, \FPC is already more powerful than \FP on unordered graphs. Still, it is not powerful enough to capture \PTIME by a result of Cai, F\"urer and Immerman~\cite{cai92optimal}. This fact will be used in the next section to prove similar negative results for specific classes of graphs. For this, we still need the notion of a \emph{graph interpretation}, which is a restricted version of the more general concept of a syntactical interpretation.

\begin{definition}\label{def:graphInterpretation}
 An \emph{$\ell$-ary graph interpretation} is a tuple $\Gamma = ( \gamma_V(\vec x), \gamma_\approx (\vec x, \vec y), \gamma_E (\vec x, \vec y) )$ of \FOL-formulas so that $|\vec x| = |\vec y| = \ell$ and in any graph, $\gamma_\approx$ defines an equivalence relation on $\gamma^G_V[\cdot]$. If $G = (V,E)$ is a graph, then $\Gamma[G] = (V_\Gamma, E_\Gamma)$ denotes the graph with vertex set $V_\Gamma = \gamma^G_V[\cdot] \modout \approx$ and edge set $E_\Gamma = \gamma^G_E [\cdot,\cdot] \modout \approx^2$.
\end{definition}

\begin{lem}[Graph Interpretations Lemma]\label{lem:graphInterpretationsLemma}
 Let $\Gamma$ be an $\ell$-ary graph interpretation. Then for any \FPC-sentence $\phi$ there is a sentence $\phi^{-\Gamma}$ with the property that $G \models \phi^{-\Gamma} \Longleftrightarrow \Gamma[G] \models \phi$.
\end{lem}

\proof[Idea.] A proof of this fact for first-order logic can be found in \cite{ebbinghaus94mathematical}. It essentially consists in modifying occurrences of the edge relation symbol and quantification in $\phi$ with the right versions of $\gamma_V$, $\gamma_\approx$ and $\gamma_E$. Lemma \ref{lem:countEqClasses} below is needed in order to deal with counting quantifiers in a sensible manner. We omit the details. \qed

\subsection{\FPC-definable canonization}

Results that prove the capturing of \PTIME on a certain graph class usually do so by showing that there is a logically definable canonization mapping from the graph structure to the number sort. Theorem \ref{thm:captureIntGraphs} will also be proved in this way, showing that there is an \FPC-formula $\varepsilon(x,y)$ with numeric variables $x$ and $y$ so that any interval graph $G = (V,E)$ is isomorphic to $\left( [|V|], \epsilon^G [\cdot,\cdot] \right)$. Since the number sort $N_V$ is linearly ordered, the Immerman-Vardi Theorem \ref{thm:immermanVardi} then implies that any \PTIME-computable property of interval graphs can be defined in \FPC.

Cai, F\"urer and Immerman have observed that for graph classes which admit \FPC-definable canonization, a generic method known as the Weisfeiler-Lehman (WL) algorithm can be used to decide graph isomorphism in polynomial time (cf. \cite{cai92optimal}). Thus by Theorem \ref{thm:captureIntGraphs}, the WL algorithm also decides isomorphism of interval graphs. In the light of efficient linear-time isomorphism algorithms for interval graphs, the novelty here lies in the fact that a simple combinatorial algorithm decides interval graph isomorphism without specifically exploiting these graphs' inherent structure. The algorithm is generic in the sense that it also decides isomorphism of planar graphs, graphs of bounded treewidth, and many others.

\subsection{Basic formulas}

We finish this section by noting some basic constructions that can be expressed in \FPC. The existence of these formulas is essentially folklore, and variants of them can for example be found in \cite{graedel07finite}. These results lay the technical foundation for a higher-level description of the canonization procedure in Section \ref{sec:CapOnIntGraphs}. We omit their straight-forward proofs.

\begin{lem}[Counting equivalence classes] \label{lem:countEqClasses}
Suppose $\sim$ is an \FPC-definable equivalence relation on $k$-tuples of $V$, and let $\phi(\vec x)$ be an \FPC-formula with $|\vec x| = k$. Assume that $\sim$ has at most $|V|$ equivalence classes. Then there is an \FPC-counting term giving the number of equivalence classes $[\vec v]$ of $\sim$ such that $G\models \phi[\vec u]$ for some $\vec u \in [\vec v]$. \qed
\end{lem}

\proof The idea is to construct the sum slicewise for each cardinality of equivalence classes first, which gives us control over the number of classes rather than the number of elements in these classes.  Let $\vec s, z, a,b$ be number variables. Define the relation $R(\vec s,\vec x)$ to hold if $\vec x$ is contained in a $\sim$-equivalence class of size $\vec s$ which contains some element making $\phi$ true. Using the fixed-point operator, it is then easy to define relation $S(\vec s, z) :\Leftrightarrow z \leq \sum_{\vec i \in [\vec s]} \# a \; \exists b \; \left( a < b \;\wedge\; b \cdot \vec i = \# \vec x\; R(\vec i,\vec x) \right)$ for $\vec s$ from $1$ to $|V|^k$. Then $\# z\; S(|V|^k, z)$ is the desired counting term. \qed

Let $\vec y$ be a tuple of numeric variables and let $\phi(\vec x,\vec y)$ be some formula. Using the ordering on the number sort and fixing $\vec x$, $\phi[\vec x,\cdot]$ can be considered a 0-1-string of truth values of length $|N_V|^{|\vec y|}$. If $\vec x$ is a tuple of elements so that the string defined by $\phi[\vec x,\cdot]$ is the lexicographically least of all such strings, then $\phi[\vec x,\cdot]$ is called the \emph{lexicographic leader}. Observe that such $\vec x$ need not be unique. Lexicographic leaders are used to break ties during the inductive definition of a graph's ordered canonical form.

\begin{lem}[Lexicographic leader] \label{lem:lexLeaderDefable}
Let $\vec x$ be a tuple of variables taking values in $V^k\times N_V^\ell$ and let $\vec y$ be a tuple of number variables. Suppose $\phi(\vec x, \vec y)$ is an \FPC-formula and $\sim$ is an \FPC-definable equivalence relation on $V^k\times N_V^\ell$. Then there is an \FPC-formula $\lambda(\vec x, \vec y)$ so that for any $\vec v \in V^k\times N_V^\ell$, $\lambda^G[\vec v, \cdot]$ is the lexicographic leader among the relations $\left\{ \phi^G[\vec u, \cdot] \bigmid \vec u \sim \vec v \right\}$. \qed
\end{lem}

\proof To start, there is a \FOL-sentence $\psi(\vec x, \vec y)$ so that $\psi[\vec u, \vec v]$ holds if and only if $\phi[\vec u,\cdot]$ is lexicographically smaller or equal to $\phi[\vec v, \cdot]$. Now $\lambda$ is given by
\begin{equation*}
 \lambda(\vec x, \vec y)  = \exists \vec z \left(\vec x \sim \vec z \,\wedge\, \phi(\vec z, \vec y) \,\wedge\, \forall \vec w (\vec w \sim \vec z \,\rightarrow\, \psi(\vec z, \vec w))\right)
\end{equation*}
\qed

Finally, we will repeatedly encounter the situation where the disjoint union of given graphs has to be defined in a canonical way. If $G_1 = ([v_1],E_1), \ldots, G_k = ([v_k],E_k)$ are (ordered) graphs in lexicographically ascending order, then we define their disjoint union $G = (V,E)$ on $V = \left[\sum_{i\in [k]} v_i \right]$ so that $G[\left[\sum_{j\in [i-1]} v_j + 1, \sum_{j\in [i]} v_j\right]]$ is order isomorphic to $G_i$ for all $i \in [k]$. It is easy to see that $G$ is uniquely well-defined, and we call it the \emph{lexicographic disjoint union} of $\{G_i\}_{i\in [k]}$. The following lemma says that lexicographic disjoint unions are \FPC-definable.

%
% \begin{definition}[Lexicographic disjoint union]
% Let $\mathcal G = \{ G_i = (V_i, E_i) \}_{i\in [k]}$ be graphs whose universes $V_i$ are initial segments $[|V_i|]$ of $\N$. Let $<$ be the lexicographic order on $\mathcal G$. Let $\pi$ be a permutation of $[k]$ so that $G_{\pi(1)}, \ldots, G_{\pi(k)}$ is in lexicographic order. Then the \emph{lexicographic disjoint union of $\mathcal G$} is the graph $G=(V,E)$ with universe $V = \left[ \sum_{i\in [k]} |V_i| \right]$ so that for each $i \in [k]$, the restriction of $G$ to $\left[\sum_{j \in [i-1]} |V_{\pi(j)}| +1, \sum_{j \in [i]} |V_{\pi(j)}| \right]$ is order isomorphic to $G_{\pi(i)}$.
% \end{definition}

\begin{lem}\label{lem:lexDisjUnionDefable}
Suppose $\sim$ is an \FPC-definable equivalence relation on $V^k\times [|V|]^\ell$ and let $\upsilon(\vec x, y)$, $\epsilon(\vec x, y,z)$ be \FPC-formulas with number variables $y,z$ defining graphs $\left( \upsilon^G[\vec v,\cdot], \epsilon^G[\vec v, \cdot,\cdot] \right)$ on the numeric sort for each $\vec v \in V^k\times [|V|]^\ell$. Furthermore, assume that $\upsilon^G[\vec v,\cdot] = \upsilon^G[\vec v',\cdot]$ whenever $\vec v \sim \vec v'$, and that $\sum_{[\vec v] \in V \modout \sim} |\upsilon^G[\vec v,\cdot]| \leq |V|$.  Then there is an \FPC-formula $\omega(y,z)$ defining on $\left[ \sum_{[\vec v] \in V \modout \sim} |\upsilon^G[\vec v,\cdot]|\right]$ the lexicographic disjoint union of the lexicographic leaders of $\sim$'s equivalence classes. \qed
\end{lem}

\proof Let $<$ be the strict weak order on $\sim$'s equivalence classes induced by the strict weak order on the classes' respective lexicographic $(\upsilon, \epsilon)$-leader. Using Lemma \ref{lem:lexLeaderDefable}, it is easy to define $<$, using elements from $V^k\times [|V|]^\ell$ to identify equivalence classes. Using the fixed point-operator, define $\omega$ inductively starting with the $<$-least elements, saving those elements $\vec v$ from equivalence classes that have already been considered in a relation $R$. In each step, find the $<$-least elements $L$ in $V^k\times [|V|]^\ell$ which are not in $R$, calculate the number $n$ of equivalence classes contained in $L$, and then expand $\omega$ by $n$ copies of $\lambda[\vec v,\cdot]$ (which is the same for any $\vec v \in L$). \qed

\section{Non-capturing results}\label{sec:noCaptureCompGraphs}

This section contains some negative results of \FPC not capturing \PTIME on a number of graph classes. In particular, this will be shown for bipartite graphs (Theorem \ref{thm:notCapOnBipGraphs}) using a simple construction and the machinery of graph interpretations (see Definition \ref{def:graphInterpretation}). Theorem \ref{thm:notCaptureCompGraphs} will then follow. We note that Theorem \ref{thm:notCapOnBipGraphs} has previously been obtained by Dawar and Richerby~\cite{dawar07power}. However, the method used here is more widely applicable and allows for stronger conclusions (see Remark \ref{rem:strongerConclusion}).

The results in this section are all based on the following theorem due to Cai, F\"urer, and Immerman~\cite{cai92optimal}.

\begin{fact} \label{fact:CFI}
 There is a \PTIME-decidable property $\mathcal P_{\mbox{\tiny CFI}}$ of graphs of degree $3$ which is not \FPC-definable.
\end{fact}

For any graph $G = (V,E)$, the \emph{incidence graph} $G^I = (V\dot\cup E,F)$ is defined by $ve \in F :\Leftrightarrow v \in V$ and $v \in e \in E$. $G^I$ is bipartite and it is straightforward to define a graph interpretation $\Gamma$ so that for any graph $G$ it holds that $\Gamma[G] \isom G^I$. Furthermore, given a graph $G^I$, it is a simple $\PTIME$-computation to uniquely reconstruct $G$ from $G^I$. Also, since the two parts of a bipartite graph can be found in linear time, it is clear how to decide whether a given graph $H$ is isomorphic to $G^I$ for some graph $G$.

\begin{thm} \label{thm:notCapOnBipGraphs}
 \FPC does not capture \PTIME on the class of bipartite graphs.
\end{thm}

\proof Recall the \PTIME-decidable query $\mathcal P_{\mbox{\tiny CFI}}$ from Fact \ref{fact:CFI} and let $\mathcal P^I := \{ H \bigmid H \isom G^I$ for some $G\in \mathcal P_{\mbox{\tiny CFI}}\}$. By the remarks above, $\mathcal P^I$ is \PTIME-decidable.

So suppose that \FPC captures \PTIME on the class of bipartite graphs. Then there is an \FPC-sentence $\phi$ such that for every bipartite graph $G$ it holds that $G\models \phi$ if and only if $G \in \mathcal P^I$. By an application of the Graph Interpretations Lemma \ref{lem:graphInterpretationsLemma} we then obtain a sentence $\phi^{-\Gamma}$ so that $G\models \phi^{-\Gamma}$ if and only if $G^I \isom \Gamma[G] \models \phi$. Thus, $\phi^{-\Gamma}$ defines $\mathcal P_{\mbox{\tiny CFI}}$, contradicting Fact \ref{fact:CFI}.\qed

Theorem \ref{thm:notCaptureCompGraphs} is now a simple corollary of the following Lemma.

\begin{lem}\label{lem:bipGraphIsCompGraph}
Every bipartite graph $G= (U\dot\cup V,E)$ is a comparability graph.
\end{lem}

\proof A suitable partial order $<$ on $U\dot\cup V$ is defined by letting $u<v$ if and only if $u \in U$, $v\in V$, and $uv \in E$.\qed

\begin{cor}
 \FPC does not capture \PTIME on the class of incomparability graphs.\qed
\end{cor}

This tells us that being a comparability or incomparability graph alone is not sufficient for a graph $G$ to be uniformly \FPC-canonizable. Section \ref{sec:CapOnIntGraphs}, however, is going to show that this is possible if $G$ is both chordal \emph{and} an incomparability graph, i.e., an interval graph. In a way, this is not simply a corollary of a capturing result on a larger class of graphs, as it is shown now that \FPC does not capture \PTIME on the class of chordal graphs, either. The construction is due to Grohe~\cite{grohe09fixed-point}.

For any graph $G = (V,E)$, the \emph{split incidence graph} $G^S = (V\dot\cup E, \binom V2 \cup F)$ is given by $ve \in F :\Leftrightarrow v \in V$ and $v \in e \in E$. Notice that $G^S$ differs from $G^I$ only by the fact that all former vertices $v\in V$ form a clique in $G^S$. %Obviously, $G^S$ is a split graph.

Given the similarity of $G^S$ and $G^I$, the analysis for split incidence graphs is completely analogous to the one for incidence graphs above. In particular, the class $\mathcal P^S := \{ H \bigmid H \isom G^S \text{ for some } G\in \mathcal P_{\mbox{\tiny CFI}}\}$ is \PTIME-decidable and given a split graph $H$, the graph $G$ for which $G^S \isom H$ can be reconstructed in \PTIME if such $G$ exists. Also, there is a graph interpretation $\Gamma'$ so that for any graph $G$: $\Gamma'[G] \isom G^S$. The proof of the following theorem is then clear, and the subsequent lemmas complete the analysis for chordal graphs.

\begin{thm}%[Grohe~\cite{grohe09fixed-point}]
 \FPC does not capture \PTIME on split graphs.\qed
\end{thm}

\begin{lem}
Every split graph $G= (U\dot\cup V,E)$ is chordal.\qed
\end{lem}

% \proof Suppose $v_1, \ldots v_k \in K\cup V$ form a chordless cycle in $G$. Since $K$ is a clique, at most $2$ of the $v_i$ can be from $K$, and they have to occur consecutively. Since $V$ is an independent set, at most half of the $v_i$ can be from $V$, and they can never occur as neighbors. It follows immediately that $k \leq 3$.\qed

\begin{cor}[Grohe~\cite{grohe09fixed-point}]
 \FPC does not capture \PTIME on the class of chordal graphs.\qed
\end{cor}

\begin{remark} \label{rem:strongerConclusion}
In fact, the proofs here admit even stronger conclusions: any \emph{regular logic} (cf. \cite{ebbinghaus94mathematical}) captures \PTIME on the class of comparability graphs (respectively chordal graphs) if and only if it captures \PTIME on the class of all graphs.
\end{remark}

Let us conclude this section by noting some non-capturing results for further intersection graph classes. A $t$-interval graph is the intersection graph of sets which are the union of $t$ intervals. By a result of Griggs and West \cite{griggs79extremal}, any graph of maximum degree 3 is a $2$-interval graph, so Fact \ref{fact:CFI} directly implies that \FPC does not capture \PTIME on $t$-interval graphs for $t \geq 2$. In \cite{Uehara08simple}, Uehara gives a construction that implies such a non-capturing result for intersection graphs of axis-parallel line segments in the plane. It follows that \FPC does not capture \PTIME on boxicity-$d$ graphs for $d \geq 2$, where a boxicity-$d$ graph is the intersection graph of axis-parallel boxes in $\mathbb R^d$ and the boxicity-$1$ graphs are just the interval graphs.

\section{Capturing \PTIME on interval graphs}\label{sec:CapOnIntGraphs}

The goal of this section is to prove Theorem \ref{thm:captureIntGraphs} by canonization. We will exhibit a numeric \FPC-formula $\varepsilon(x,y)$ so that for any interval graph $G = (V,E)$, $([|V|], \varepsilon^G[\cdot])$ defines a graph on the numeric sort of \FPC which is isomorphic to $G$. The canonization essentially consists of finding the lexicographic leader among all possible interval representations of $G$. For this, as discussed above, it is enough to bring the maximal cliques of $G$ in the right linear order. The first lemma shows that the maximal cliques of $G$ are \FOL-definable.

\begin{lem} \label{lem:maxCliquesDefable}
Let $G = (V,E)$ be an interval graph and let $M$ be a maximal clique of $G$. Then there are vertices $u,v \in M$, not necessarily distinct, such that $M = N(u)\cap N(v)$.
\end{lem}

This is fairly intuitive. Consider the following minimal interval representation of a graph $G$.

%\begin{figure}[htb]
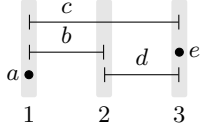
\begin{center}
\begin{tikzpicture}
 \foreach \x in {1,...,3}
{
 \fill[xshift=\x cm-1cm,rounded corners=1pt,fill=gray!20!white] (-3pt,-.3cm) rectangle (3pt,1cm);
 \draw[xshift=\x cm-1cm] (0,-.3cm) node[anchor=north] {\small $\x$};
}

 \filldraw (0,0) node[anchor=east] {\small $a$} circle (1.5pt)
	(2,.3) node[anchor=west] {\small $e$} circle (1.5pt);
 \draw[|-|] (0,.3) -- (1,.3) node[midway,above] {\small $b$};
 \draw[|-|] (0,.7) -- (2,.7) node[near start,above] {\small $c$};
 \draw[|-|] (1,0) -- (2,0) node[midway,above] {\small $d$};
\end{tikzpicture}
\end{center}
%\caption{An interval graph.}\label{fig:maxCliqDefTwoVertices}
%\end{figure}

Max cliques $1$ and $3$ are precisely the neighborhoods of vertices $a$ and $e$, respectively. The vertex pairs $(a,a)$, $(a,b)$, and $(a,c)$ all define max clique $1$, and similarly three different vertex pairs define max clique $3$. Max clique $2$ is not the neighborhood of any single vertex, but it is uniquely defined by $N(b)\cap N(d)$.

\proof[Proof of Lemma \ref{lem:maxCliquesDefable}]  Let $\mathcal I$ be a minimal interval representation of $G$. First assume that $M$ is the $\lhd_{\mathcal I}$-least maximal clique. The lemma is trivial if $M$ is the only maximal clique of $G$, otherwise let $X$ be $M$'s $\lhd_{\mathcal I}$-successor. Since $M \neq X$ there is $v \in M\setminus X$, and $M$ is the only maximal clique of $G$ that $v$ is contained in (as $v$ is contained in $\lhd_{\mathcal I}$-consecutive max cliques). Hence, $M = N(v)$. A symmetric argument holds if $M$ is the $\lhd_{\mathcal I}$-greatest maximal clique. Now assume that $M$ is neither $\lhd_{\mathcal I}$-least nor maximal and let $X, Y$ be $M$'s immediate $\lhd_{\mathcal I}$-predecessor and successor, respectively. There exist $x \in M\setminus X$ and $y \in M\setminus Y$, and we claim that $N(x) \cap N(y) = M$. In fact, since any vertex in $M$ is contained both in $N(x)$ and $N(y)$, we have $M \subseteq N(x) \cap N(y)$. Now let $u \in N(x) \cap N(y)$ and write $I_u = [a,b]$. Let $k$ be the (unique) integer such that $M(k) = M$. Then $ux \in E$ implies $b \geq k$, and $vx \in E$ implies $a \leq k$. Thus, $k \in I_u$ and hence $u \in M$, which proves the claim. \qed

% \begin{cor}
% There are \FOL-formulas $\kappa(x,y)$ and $\epsilon(x_1,y_1,x_2,y_2)$ with the following properties:
% \begin{itemize}
% \item $G \models \kappa[u,v]$ if and only if $N(u)\cap N(v)$ is a maximal clique of $G$,
% \item $G \models \epsilon[u_1, v_1, u_2, v_2]$ if and only if $(u_1,v_1)$ and $(u_2,v_2)$ define the same maximal clique of $G$. \qed
% \end{itemize}
% \end{cor}

Now, whether or not a vertex pair $(u,v) \in V^2$ defines a max clique is easily definable in \FOL, as is the equivalence relation on $V^2$ of vertex pairs defining the same max clique. Lemma \ref{lem:maxCliquesDefable} tells us that \emph{all} max cliques can be defined by such vertex pairs. For any $v \in V$, let the \emph{span of $v$}, denoted $\spn(v)$, be the number of max cliques of $G$ that $v$ is contained in. Since equivalence classes can be counted by Lemma \ref{lem:countEqClasses}, $\spn(x)$ is \FPC-definable on the class of interval graphs by a counting term with $x$ as a free vertex variable.

Generally representing max cliques by pairs of variables $(x,y) \in V^2$ allows us to treat max cliques as first-class objects that can be quantified over. For reasons of conceptual simplicity, the syntactic overhead which is necessary for working with this representation will not be made explicit in the remainder of this section.

\subsection{Extracting information about the order of the maximal cliques}
Now that we are able to handle maximal cliques, we would like to simply pick an end of the interval graph $G$ and work with the order which this choice induces on the rest of the maximal cliques. Of course, the choice of an end does not necessarily impose a linear order on the maximal cliques. However, the following recursive procedure turns out to recover all the information about the order of the max cliques induced by choosing an end of $G$.

Let $\mathcal M$ be the set of maximal cliques of an interval graph $G = (V,E)$ and let $M \in \mathcal M$. The binary relation $\prec_M$ is defined recursively on the elements of $\mathcal M$ as follows:
\begin{align*}
\mbox{Initialization:}& \quad M \prec_M C \mbox{ for all } C \in \mathcal M \setminus \{M\}\\
 %C\prec_M D & \quad\mbox{if } \begin{cases} \exists v \in C\setminus D \mbox{ and } \exists E \in \mathcal M \mbox{ with } \\ \qquad E \prec_M D \mbox{ and } v \in E, \mbox{ or}\\
%  \exists v \in D\setminus C \mbox{ and } \exists E\in \mathcal M \mbox{ with } \\ \qquad C \prec_M E \mbox{ and } v \in E. \tag{$\bigstar$} \end{cases}\label{eqn:indDef}
  C\prec_M D & \quad\mbox{if } \begin{cases} \exists E \in \mathcal M \mbox{ with } E \prec_M D \mbox{ and } (E\cap C) \setminus D \neq \emptyset \quad\mbox{or}\\
   \exists E \in \mathcal M \mbox{ with } C \prec_M E \mbox{ and } (E\cap D) \setminus C \neq \emptyset. \tag{$\bigstar$} \end{cases}\label{eqn:indDef}
\end{align*}

The following interval representation of a graph $G$ illustrates this definition.

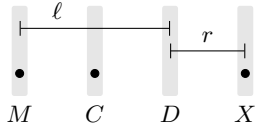
\begin{center}
\begin{tikzpicture}
 \foreach \x/\xtext in {1/M,2/C,3/D,4/X}
{
 \fill[xshift=\x cm-1cm,rounded corners=1pt,fill=gray!20!white] (-3pt,-.3cm) rectangle (3pt,.9cm);
 %\draw[xshift=\x cm-1cm] (0,-.3cm) node[anchor=north] {\small $\x$};
 \draw[xshift=\x cm-1cm] (0,-.3cm) node[anchor=north] {\small $\xtext$};
}

 \filldraw (0,0) circle (1.5pt)
	(1,0) circle (1.5pt)
	(3,0) circle (1.5pt);
 %\draw[|-|,] (0,1) -- (1,1) node[midway,above] {\small $\ell$};
 \draw[|-|] (0,.6) -- (2,.6) node[near start,above] {\small $\ell$};
 %\draw[|-|] (1,.5) -- (2,.5);
 \draw[|-|] (2,.3) -- (3,.3) node[midway,above] {\small $r$};	

\end{tikzpicture}
\end{center}

Suppose we have picked max clique $M$, then $C \prec_M X$ and $D \prec_M X$ since $\ell \in C\cap D\cap M \setminus X$ and $M \prec_M X$ by the initialization step. In a second step, it is determined that $C \prec_M D$ since $r\in D\cap X \setminus C$ and $C \prec_M X$. So in this example, $\prec_M$ actually turns out to be a strict linear order on the max cliques of $G$. This is not the case in general, but $\prec_M$ will still be useful when $M$ is a possible end of $G$. The definition of $\prec_M$ seems natural to me for the task of ordering the max cliques of an interval graphs. However, I am not aware of it appearing previously anywhere in the literature.

It is readily seen how to define $\prec_M$ using the inflationary fixed-point operator, where maximal cliques are defined by pairs of vertices from $G$.

\begin{remark} \label{rem:alreadySTC}
 In fact, $\prec_M$ can already be defined using a \emph{symmetric transitive closure} operator as follows: define an edge relation on $\mathcal M^2$ by connecting $(C,D)$ and $(E,F)$ if $E\prec_M F$ follows from $C \prec_M D$ by one application of (\ref{eqn:indDef}). Inspection of (\ref{eqn:indDef}) shows that this edge relation is symmetric, hence the graph is undirected. Now $C \prec_M D$ holds if and only if $(C,D)$ is reachable from $(M,X)$ for some max clique $X$. This observation is used in \cite{koebler10interval} to show that canonical forms of interval graphs can be computed using only logarithmic space.
\end{remark}

The following lemmas prove important properties of $\prec_M$. We say that a binary relation $R$ on a set $A$ is \emph{asymmetric}\index{asymmetric relation} if $ab \in R$ implies $ba \not\in R$ for all $a,b\in A$. In particular, asymmetric relations are irreflexive.

\begin{lem} \label{lem:antisymImpTransitive}
If $\prec_M$ is asymmetric, then it is transitive. Thus, if $\prec_M$ is asymmetric, then it is a strict partial order.
\end{lem}

\proof By a derivation chain of length $k$ we mean a finite sequence $X_0 \prec_M Y_0$, $X_1 \prec_M Y_1$, $\ldots$, $X_k \prec_M Y_k$ such that $X_0 = M$ and for each $i \in [k]$, the relation $X_i \prec_M Y_i$ follows from $X_{i-1} \prec_M Y_{i-1}$ by one application of (\ref{eqn:indDef}). Clearly, whenever it holds that $X \prec_M Y$ there is a derivation chain that has $X \prec_M Y$ as its last element.

So assume that $\prec_M$ is asymmetric. Suppose $A \prec_M B \prec_M C$ and let a derivation chain $(L_0, \ldots, L_a)$ of length $a$ be given for $A \prec_M B$. The proof is by induction on $a$. If $a=0$, then $A = M$ and $A \prec_M C$ holds. For the inductive step, suppose $a = n$ and consider the second to last element $L_{n-1}$ in the derivation chain. There are two cases:
\begin{itemize}
\item $L_{n-1} = (X \prec_M B)$ and there is a vertex $v \in (X\cap A)\setminus B$: By induction it holds that $X \prec_M C$. Now if we had $v \in C$, the fact that $A \prec_M B$ would imply $C \prec_M B$, which contradicts asymmetry of $\prec_M$. Hence, $v\not\in C$ and one more application of (\ref{eqn:indDef}) yields $A \prec_M C$.
\item $L_{n-1} = (A \prec_M X)$ and there is a vertex $v \in (X\cap B)\setminus A$: If $v \in C$, then we immediately get $A \prec_M C$. If $v\not\in C$, then $X \prec_M C$. Thus we can derive $A \prec_M X \prec_M C$ where the left derivation chain has length $n-1$. By induction, $A \prec_M C$ follows.\qed
\end{itemize}

\begin{lem}\label{lem:moduleImpIncomparable}
Let $\mathcal C \subset \mathcal M$ be a set of max cliques with $M \not\in \mathcal C$. Suppose that for all $A\in \mathcal M\setminus\mathcal C$ and any $C,C'\in\mathcal C$ it holds that $A\cap C = A\cap C'$. Then the max cliques in $\mathcal C$ are mutually incomparable with respect to $\prec_M$.
\end{lem}

\proof Suppose for contradiction that there are $C,C'\in\mathcal C$ with $C\prec_M C'$. Let $M \prec_M Y_0$, $X_1 \prec_M Y_1$, $\ldots$, $X_k \prec_M Y_k$ be a derivation chain for $C\prec_M C'$ as in the proof of Lemma \ref{lem:antisymImpTransitive}. Since $X_k = C$, $Y_k = C'$, and $M\not\in \mathcal C$, there is a largest index $i$ so that either $X_i$ or $Y_i$ is not contained in $\mathcal C$.

If $X_i \not\in \mathcal C$, then $X_{i+1}\in\mathcal C$ and $Y_i=Y_{i+1} \in\mathcal C$ and it holds that $X_i \cap X_{i+1}\setminus Y_{i+1} \neq \emptyset$. Consequently, $X_i \cap X_{i+1} \neq X_i \cap Y_{i+1}$, contradicting the assumption of the lemma. Similarly, if $Y_i \not\in \mathcal C$, then $Y_{i+1}\in\mathcal C$ and $X_i = X_{i+1}\in\mathcal C$ and it holds that $Y_i\cap Y_{i+1}\setminus X_{i+1} \neq \emptyset$. Thus, $Y_i \cap Y_{i+1} \neq Y_i \cap X_{i+1}$, again a contradiction.\qed

In fact, there is a converse to Lemma \ref{lem:moduleImpIncomparable} when the set of $\prec_M$-incomparable max cliques is maximal.

\begin{lem} \label{lem:sameIntersection}
Suppose $M$ is a max clique of $G$ % so that $\prec_M$ is a strict partial order and
and $\mathcal C$ is a maximal set of $\prec_M$-incomparable max cliques. Let $D \in\mathcal M\setminus\mathcal C$. Then $D\cap C = D\cap C'$ for all $C,C' \in \mathcal C$.
\end{lem}

\proof We say that a max clique $A$ \emph{splits} a set of max cliques $\mathcal X$ if there are $X,Y \in \mathcal X$ so that $A\cap X \neq A\cap Y$. %$A \cap X \setminus Y \neq \emptyset$.
If in addition to splitting $\mathcal X$, $A$ is also $\prec_M$-comparable to all the elements in $\mathcal X$, then either $A\cap X\setminus Y\neq \emptyset$ or $A\cap Y\setminus X \neq \emptyset$ and one application of (\ref{eqn:indDef}) implies that $X$ and $Y$ are comparable.

Suppose for contradiction that there is $X_1 \in\mathcal M\setminus\mathcal C$ splitting $\mathcal C$.
%with $X_1\cap C \setminus C' \neq \emptyset$ for some $C,C'\in \mathcal C$. We say that $X_1$ \emph{splits} $\mathcal C$. Choose some $x_1\in X_1\cap C\setminus C'$.
We greedily grow a list of max cliques $X_i$ with the property that $X_i \in \mathcal M \setminus (\mathcal C \cup \{X_1,\ldots,X_{i-1}\})$ splits the set $\mathcal X_{i-1} := \mathcal C \cup \{X_1,\ldots,X_{i-1}\}$.
%So $X_i \cap A \setminus B \neq \emptyset$ for some $A,B \in \mathcal X_{i-1}$ and we choose corresponding vertices $x_i \in X_i \cap A \setminus B$.
The list $X_1, \ldots, X_k$ is complete when no further max clique splits the set $\mathcal X_k$.

Suppose that $M \not\in \mathcal X_k$. For any $D\in\mathcal M\setminus\mathcal X_k$ we have %$\emptyset = D \cap X \setminus X' = (D\cap X) \setminus (D\cap X')$ for all $X,X' \in \mathcal X_k$, hence
$D\cap X = D\cap X'$ for all $X,X'\in\mathcal X_k$, so Lemma \ref{lem:moduleImpIncomparable} implies that the max cliques in $\mathcal X_k$ are $\prec_M$-incomparable. However, this is impossible since we assumed $\mathcal C \subsetneq \mathcal X_k$ to be maximal. Therefore, $M\in\mathcal X_k$.

%Next, we claim that whenever a max clique $S$ splits a set $\mathcal Z$ ($S\not\in\mathcal Z$) but not $\mathcal Z \setminus \{X\}$ for some $X\in\mathcal Z$, then $S$ splits $\{X,Y\}$ for all $Y\in \mathcal Z \setminus \{X\}$. There are two cases to consider. Either, there is $v\in S\cap X \setminus A$ for some $A\in \mathcal Z$ and $S\cap Y \setminus A = \emptyset$ for all $Y\in \mathcal Z \setminus \{X\}$. Then $v \not\in Y$, so $v \in S\cap X \setminus Y$ and $S$ splits $\{X,Y\}$. Or, there is $v\in S\cap A \setminus X$ for some $A\in \mathcal Z$ and $S\cap A \setminus Y = \emptyset$
%

Now let $Y_{1},\ldots Y_{\ell}$ be a shortest list of max cliques from $\mathcal X_k$ so that $Y_{\ell} = M$ and each $Y_{j}$ splits $\mathcal Y_{j-1} := \mathcal C \cup \{Y_{1},\ldots,Y_{{j-1}}\}$.

\begin{claim}\label{claim:auxLemSameIntersection}
For all $j\in [2,\ell]$, $Y_j\cap Y_{j-1} \neq Y_j \cap A$ for all $A\in \mathcal Y_{j-2}$.
\end{claim}
\proof[Proof of Claim \ref{claim:auxLemSameIntersection}] Consider $j = \ell$ and suppose that there is $A\in \mathcal Y_{\ell -2}$ with $Y_\ell \cap A = Y_\ell \cap Y_{\ell - 1}$. As $Y_\ell$ splits $\mathcal Y_{\ell-1}$, there must be some $B\in \mathcal Y_{\ell-2}$ such that $Y_\ell \cap B \neq Y_\ell \cap Y_{\ell - 1}$. But then $Y_\ell$ already splits $\mathcal Y_{\ell -2}$, so by eliminating $Y_{\ell - 1}$ we could make the list shorter.

Inductively, suppose that the claim holds for all $i > j$, but not for $j$. Then there are $A,B \in \mathcal Y_{j-2}$ such that $Y_j \cap B \neq Y_j \cap Y_{j-1} = Y_j \cap A$, so $Y_j$ already splits $\mathcal Y_{j-2}$. Removing $Y_{j-1}$ from the list gives us a shorter list in which $Y_i$ still splits $\mathcal Y_{i-1}$ for all $i> j$ because of our inductive assumption. As we assumed our list to be shortest, this concludes the inductive step. \qedd{Claim \ref{claim:auxLemSameIntersection}}

We now argue once again inductively backwards down the list $Y_{1}, \ldots, Y_{\ell}$ with the goal of showing that $Y_{j}$ is $\prec_M$-comparable to all max cliques in $\mathcal Y_{{j-1}}$. Certainly, this is true for $Y_{\ell} = M$ and $\mathcal Y_{\ell - 1}$. Assume that $Y_{j}$ is comparable to all max cliques in $\mathcal Y_{{j-1}}$ for $j\in [2,\ell]$. Since $Y_j \cap Y_{j-1} \neq Y_j \cap A$ for all $A\in \mathcal Y_{j-2}$ by Claim \ref{claim:auxLemSameIntersection}, it follows that $Y_{{j-1}}$ is comparable to all max cliques in $\mathcal Y_{j-2}$.

Now $Y_{1}$ is comparable to all max cliques in $\mathcal C$. Since $Y_{1}$ splits $\mathcal C$, there are $C,C'\in \mathcal C$ so that $C\prec_M C'$, contradicting our assumption that the max cliques in $\mathcal C$ are $\prec_M$-incomparable. Therefore we conclude that there is no $D\in \mathcal M\setminus \mathcal C$ splitting $\mathcal C$.\qed

% \begin{lem} \label{lem:sameIntersection}
% Suppose $M$ is a max clique of $G$ so that $\prec_M$ is a strict partial order and $\mathcal C$ is a maximal set of incomparable max cliques. Let $D \in\mathcal M\setminus\mathcal C$. Then $D\cap C = D\cap C'$ for all $C,C' \in \mathcal C$.
% \end{lem}
% 
% \proof Suppose there are $C,C' \in \mathcal C$ and $D \in \mathcal M \setminus \mathcal C$ with $C \cap D \neq C' \cap D$. Without loss of generality assume that there is an element $x \in (C\cap D) \setminus C'$ (otherwise, swap $C$ for $C'$). As $\mathcal C$ is maximal, there is some $C_D \in \mathcal C$ such that $C_D \prec_M D$ or $D \prec_M C_D$. Without loss of generality assume the former. Now if $x \not\in C_D$, then (\ref{eqn:indDef}) implies $C_D \prec_M C$, which contradicts the assumption that the elements of $\mathcal C$ are mutually incomparable. But if $x \in C_D$, then (\ref{eqn:indDef}) implies that $C' \prec_M C_D$ since $x\not\in C'$, again contradicting incomparability. \qed

Lemma \ref{lem:sameIntersection} says that %if we are lucky enough to choose a max clique $M$ of $G$ so that $\prec_M$ becomes a strict partial order, then 
incomparable max cliques interact with the rest of $\mathcal M$ in a uniform way. Let us make this notion more precise. A \emph{module} of $G$ is a set $S \subseteq V$ so that for any vertex $x \in V\setminus S$, $S$ is either completely connected or completely disconnected to $x$. In other words, for all $u,v \in S$ and all $x \in V\setminus S$ it holds that $ux \in E \leftrightarrow vx \in E$. The next drawing illustrates the occurrence of a module in an interval graph.

%\vspace{-.4cm}
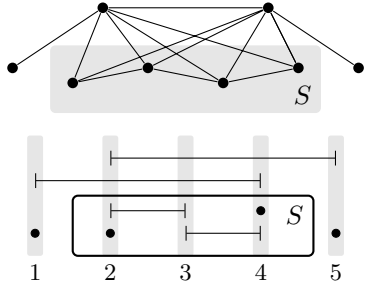
\begin{center}
\begin{tikzpicture}[v/.style={circle,fill,inner sep=0pt,minimum size=4pt}]
 \foreach \x in {1,...,5}
{
 \fill[xshift=\x cm-1cm,rounded corners=1pt,fill=gray!20!white] (-3pt,-.3cm) rectangle (3pt,1.3cm);
 %\draw[xshift=\x cm-1cm] (0,-.3cm) node[anchor=north] {\small $\x$};
 \draw[xshift=\x cm-1cm] (0,-.3cm) node[anchor=north] {\small $\x$};
}

 \draw[thick,rounded corners=2pt] (.5,-.3cm) rectangle (3.7,.5) node[anchor=north east] {$S$};

 \filldraw (0,0) circle (1.5pt)
	(1,0) circle (1.5pt)
	(3,.3) circle (1.5pt)
	(4,0) circle (1.5pt);

 \draw[|-|] (0,.7) -- (3,.7);
 \draw[|-|] (1,1) -- (4,1);
 \draw[|-|] (1,.3) -- (2,.3);
 \draw[|-|] (2,0) -- (3,0);

 \def\x{-1}
 \def\y{2};
 \fill[gray!20!white,rounded corners=2pt] (\x+1.2,\y+.5) rectangle (\x+4.8,\y-.4) node[anchor=south east,black] {$S$};

 \node[v] (1) at (\x+.7,\y+.2) {};
 \node[v] (2) at (\x+1.9,\y+1) {};
 \node[v] (3) at (\x+1.5,\y) {};
 \node[v] (4) at (\x+2.5,\y+.2) {};
 \node[v] (5) at (\x+3.5,\y) {};
 \node[v] (6) at (\x+4.5,\y+.2) {};
 \node[v] (7) at (\x+4.1,\y+1) {};
 \node[v] (8) at (\x+5.3,\y+.2) {};

\draw (1) -- (2) -- (3) -- (4) -- (5) -- (6) -- (7) -- (8);
\draw (2) -- (4) -- (7) -- (5) -- (2) -- (6) -- (7) -- (3);
\draw (2) -- (7);

\end{tikzpicture}
\end{center}

%Recall that the set of max cliques is denoted $\mathcal M$.

\begin{cor}\label{cor:incompImpModule}
Suppose $M$ is a max clique of $G$ so that $\prec_M$ is a strict partial order and $\mathcal C$ is a maximal set of incomparable max cliques. Then
\begin{itemize}
 \item $S_{\mathcal C} := \bigcup_{C \in \mathcal C} C \setminus \bigcup_{D \in \mathcal M \setminus \mathcal C}D$ is a module of $G$, and
 \item $S_{\mathcal C} = \left\{ v \in \bigcup\mathcal C \;\big|\; \spn(v) \leq |\mathcal C| \right\}$.
\end{itemize}
\end{cor}

\proof Let $u,v \in S_{\mathcal C}$ and $x \in V\setminus S_{\mathcal C}$ and suppose that $ux\in E$, but $vx \not\in E$. There is a max clique $C \in \mathcal M$ with $u,x \in C$, but $v \not\in C$, and since $u \in S$ we must have $C \in\mathcal C$. By the definition of $S_{\mathcal C}$, $x$ is also contained in some max clique $D \in \mathcal M\setminus \mathcal C$. Finally, let $C'$ be some max clique in $\mathcal C$ containing $v$, so $x \not\in C'$. Thus, $D \cap C \neq D \cap C'$, contradicting Lemma \ref{lem:sameIntersection}.

For the second statement, let $v \in \bigcup\mathcal C$. If $v \in S_{\mathcal C}$, then clearly $\spn(v) \leq |\mathcal C|$. But if $v \not\in S_{\mathcal C}$, then it is contained in some $D \in\mathcal M\setminus\mathcal C$, and by Lemma \ref{lem:sameIntersection} $v$ must also be contained in all max cliques in $\mathcal C$. Thus, $\spn(v) > |\mathcal C|$, proving the statement. \qed

This characterization of the modules occurring when defining the relations $\prec_M$ will be central in the canonization procedure of $G$. There is another corollary of Lemma \ref{lem:sameIntersection} which proves that $\prec_M$ has a particularly nice structure.

\begin{cor}\label{cor:strictOrderImpWeakOrder}
 If $M$ is a max clique of $G$ so that $\prec_M$ is a strict partial order, then $\prec_M$ is a strict weak order.
\end{cor}

\proof We need to prove that $\prec_M$-incomparability is a transitive relation of $G$'s max cliques. So let $(A,B)$ and $(B,C)$ be incomparable pairs with respect to $\prec_M$. Let $\mathcal C_{AB}$ and $\mathcal C_{BC}$ be maximal sets of incomparables containing $\{A,B\}$ and $\{B,C\}$, respectively. By Lemma \ref{lem:sameIntersection}, we have $D\cap X = D\cap B = D\cap Y$ for every $X,Y\in \mathcal C_{AB} \cup \mathcal C_{BC}$ and $D\in \mathcal M \setminus (\mathcal C_{AB} \cup \mathcal C_{BC})$. As $M \not\in \mathcal C_{AB} \cup \mathcal C_{BC}$ Lemma \ref{lem:moduleImpIncomparable} implies that the max cliques in $\mathcal C_{AB} \cup \mathcal C_{BC}$ are $\prec_M$-incomparable, so in particular $A$ and $C$ are incomparable with respect to $\prec_M$.\qed

% \proof We need to prove that incomparability is a transitive relation of max cliques of $G$. So let $(A,B)$ and $(B,C)$ be incomparable pairs with respect to $\prec_M$. Then by (\ref{eqn:indDef}) the following holds:
% \begin{align*}
% &\neg\exists v\in A\setminus B:\; \exists F\in \mathcal M \mbox{ comparable to } A \mbox{ and } v\in F \mbox{, and}\\
% &\neg\exists v\in C\setminus B:\; \exists F\in \mathcal M \mbox{ comparable to } C \mbox{ and } v\in F
% \end{align*}
% Suppose for contradiction that $A$ and $C$ were comparable, and let $\mathcal C_{AB}$ and $\mathcal C_{BC}$ be maximal set of incomparables containing $\{A,B\}$ and $\{B,C\}$, respectively. Since $A \not\in \mathcal C_{BC}$ and $C \not\in \mathcal C_{AB}$, Lemma \ref{lem:sameIntersection} implies that $A\cap B = A\cap C = C \cap B$. But then $A\setminus B = A \setminus C$ and $C \setminus B = C \setminus A$, and the two statements above become
% \begin{align*}
% &\neg\exists v\in A\setminus C:\; \exists F\in \mathcal M \mbox{ comparable to } A \mbox{ and } v\in F \mbox{, and}\\
% &\neg\exists v\in C\setminus A:\; \exists F\in \mathcal M \mbox{ comparable to } C \mbox{ and } v\in F.
% \end{align*}
% Thus, $A$ and $C$ are incomparable, contradicting the above assumption and proving the corollary.\qed

At this point, let us put the pieces together and show that picking an arbitrary max clique $M$ as an end of $G$ and defining $\prec_M$ is a useful way to obtain information about the structure of $G$.

\begin{lem} \label{lem:partialOrderEqEnd}
Let $M$ be a max clique of an interval graph $G$. Then $\prec_M$ is a strict weak order if and only if $M$ is a possible end of $G$.
\end{lem}

\proof If $M$ is a possible end of $G$, then let $\mathcal I$ be a minimal interval representation of $G$ which has $M$ as its first clique. Let $\lhd_{\mathcal I}$ be the linear order $\mathcal I$ induces on the max cliques of $G$. In order to show asymmetry it is enough to observe that, as relations, we have $\prec_M \subseteq \lhd_{\mathcal I}$. It is readily verified that this holds true of the initialization step in the recursive definition of $\prec_M$, and that whenever max cliques $C,D$ satisfy (\ref{eqn:indDef}) with $\prec_M$ replaced by $\lhd_{\mathcal I}$, then it must hold that $C \lhd_{\mathcal I} D$. This shows asymmetry, and by Lemma \ref{lem:antisymImpTransitive} and Corollary \ref{cor:strictOrderImpWeakOrder} $\prec_M$ is a strict weak order.

Conversely, suppose $\prec_M$ is a strict weak order. The first aim is to turn $\prec_M$ into a linear order. Let $\mathcal C$ be a maximal set of incomparable max cliques, and recall the set $S_{\mathcal C} = \bigcup_{C \in \mathcal C} C \setminus \bigcup_{D \in \mathcal M \setminus \mathcal C}D$. Since $G[S_{\mathcal C}]$ is an interval graph, we can pick an interval representation $\mathcal I_{S_{\mathcal C}}$ for $G[S_{\mathcal C}]$. The set of max cliques of $G[S_{\mathcal C}]$ is given by $\left\{ C \cap S_{\mathcal C} \;\big|\; C\in\mathcal C\right\}$, and since $S_{\mathcal C}$ is a module, $C\cap S_{\mathcal C} \neq C' \cap S_{\mathcal C}$ for any $C\neq C'$ from $\mathcal C$. Thus, $\mathcal I_{S_{\mathcal C}}$ induces a linear order $\lhd_{\mathcal C}$ on the elements of $\mathcal C$. Now let $C \lhd_M D$ if and only if $C \prec_M D$, or $C, D \in \mathcal C$ for some maximal set of incomparables $\mathcal C$ and $C \lhd_{\mathcal C} D$. This is a strict linear order since $\prec_M$ is a strict weak order. We claim that $\lhd_M$ is an ordering of the max cliques which is isomorphic to the linear order induced by some interval representation of $G$. This will imply that $M$ is a possible end of $G$.

In order to prove the claim, it is enough to show that each vertex $v\in V$ is contained in consecutive max cliques. Suppose for contradiction that there are max cliques $A \lhd_M B \lhd_M C$ and $v$ is contained in $A$ and $C$, but not in $B$. Certainly, this cannot be the case if $A,B,C$ are incomparable with respect to $\prec_M$, so assume without loss of generality that $A \prec_M B$. Now, since $v \in (A\cap C)\setminus B$, (\ref{eqn:indDef}) implies that $C \prec_M B$, which contradicts the asymmetry of $\lhd_M$. \qed

\begin{remark} \label{remark:doesntWorkGenerally}
 The recursive definition of $\prec_M$ and Lemma \ref{lem:antisymImpTransitive} through Corollary \ref{cor:strictOrderImpWeakOrder} do not depend on $G$ being an interval graph. However, the proof of Lemma \ref{lem:partialOrderEqEnd} shows that $\prec_M$ only turns out to be a partial order if the max cliques can be brought into a linear order, modulo the occurrence of modules. In particular, defining $\prec_M$ in a general chordal graph does not yield any useful information if the graph's tree decomposition into max cliques requires a tree vertex of degree 3 or more, which is the case for all chordal graphs which are not interval graphs.
\end{remark}

\subsection{Canonizing when $\prec_M$ is a linear order}\label{subsec:canLinOrder}

Since $\prec_M$ is \FP-definable for any max clique $M$, and since asymmetry of $\prec_M$ is \FOL-definable, Lemma \ref{lem:partialOrderEqEnd} gives us a way to define possible ends of interval graphs in \FP. Moreover, if $M$ is a possible end of $G = (V,E)$, then $\prec_M$ contains precisely the ordering imposed on the max cliques of $G$ by the choice of $M$ as the first clique.

First, suppose that $G = (V,E)$ is an interval graph and $\prec$ is a linear order on the max cliques which is induced by an interval representation of $G$. Define the binary relation $<^G$ on the vertices of $G$ as follows. For $x \in V$, let $A_{x}$ be the $\prec$-least max clique of $G$ containing $x$. Then let
\begin{equation*}
 x <^G y :\Leftrightarrow \begin{cases} A_{x} \prec A_{y}, \mbox{ or}\\
     A_{x} = A_{y} \mbox{ and } \spn(x) < \spn(y).
                                \end{cases}
\end{equation*}

It is readily verified that $<^G$ is a strict weak order on $V$, and if $x,y$ are incomparable, then $N(x) = N(y)$. Now it is easy to canonize $G$: if $[v]$ denotes the equivalence class of vertices incomparable to $v$, then $[v]$ is represented by the numbers from the interval $[a+1 ,a + |[v]|]$, where $a$ is the number of vertices which are strictly $<^G$-smaller than $v$. Since all vertices in $[v]$ have precisely the same neighbors in $G\setminus [v]$ and $[v]$ forms a clique, it is also clear how to define the edge relation on the number sort.

Now if $G$ is any interval graph and $M$ is a possible end, we can still define an ordering for those vertices that are not contained in a module. Let $\sim^G_M$ be the equivalence relation on $V$ for which $x \sim^G_M y$ if and only if $x=y$ or there is a nonsingular maximal set of incomparables $\mathcal C$ with respect to $\prec_M$ so that $x,y \in S_{\mathcal C}$. Denote the equivalence class of $x \in V$ under $\sim^G_M$ by $[x]$, and define the edge relation $E_M$ of the graph $G_M = (V \modout \sim^G_M, E_M)$ by $[u][v] \in E_M :\Leftrightarrow \exists x\in [u], y\in [v]$ s.t. $xy \in E$. It follows directly from the definition of $\sim^G_M$ that if $A$ is a max clique which is $\prec_M$-comparable to all other max cliques in $G$, then all $v \in A$ are in singleton equivalence classes $[v] = \{v\}$. If $\mathcal C$ is a nonsingular maximal set of incomparables, then there is precisely one max clique $C$ in $G_M$ which contains all the equivalence classes associated with $\mathcal C$, i.e., $C = \left\{ [v] \bigmid v \in \bigcup \mathcal C \right\}$. Thus $\prec_M$ induces a strict linear order on the max cliques of $G_M$. In fact, this shows that $G_M$ is an interval graph with a valid interval representation induced by $\prec_M$. %Notice that $\sim^G_M$, and the graph $G_M$ are \FPC-definable.

\subsection{Canonizing general interval graphs} \label{subsec:generalIntGraphs}
What is left is to deal with the sets $S_{\mathcal C}$ coming from maximal sets of incomparables. Let $P' = \left\{ (M,n) \bigmid M \in \mathcal M, n \in [|V|] \right\}$. For each $(M,n) \in P'$ define $V_{M,n}$ as the set of vertices of the connected component of $V \setminus \left\{ v \in V \bigmid \spn(v) > n \right\}$ which intersects $M$ (if non-empty). Notice that $M_n := M \cap V_{M,n}$ is a max clique of $G[V_{M,n}]$. Finally, let $P$ be the set of those $(M,n) \in P'$ for which defining $\prec_{M_n}$ in $G[V_{M,n}]$ yields a strict partial order of $G[V_{M,n}]$'s max cliques.

It is immediate from Corollary \ref{cor:incompImpModule} that for any maximal set of incomparable max cliques $\mathcal C$, $S_{\mathcal C} = \bigcup_{C\in\mathcal C} V_{C,|\mathcal C|}$. In this situation, for any $C\in \mathcal C$, the set $V_{C,|\mathcal C|}$ defines a component of $S_{\mathcal C}$, and $(C,|\mathcal C|) \in P$ if and only if $C \cap S_{\mathcal C}$ is a possible end of (one of the components of) $G[S_{\mathcal C}]$. This gives us enough structure to perform canonization.

\proof[Proof of Theorem \ref{thm:captureIntGraphs}] We define the relation $\epsilon(M,n,x,y)$ inductively, where $(M,n) \in P$ and $x,y$ are number variables. $([|V_{M,n}|], \epsilon^G[M,n,\cdot,\cdot])$ will be an isomorphic copy of $G[V_{M,n}]$ on the numeric sort. To this end, start defining $\epsilon$ for all $(M,1) \in P$, then for all $(M,2) \in P$, and so on, up to all $(M,|V|) \in P$.

Suppose we want to define $\epsilon$ for $(M,n) \in P$, then first compute the strict weak order $\prec_{M_n}$ on the interval graph $G[V_{M,n}]$. Consider any nonsingular maximal set of incomparables $\mathcal C$ and let $m := |\mathcal C|$. Let $H_1, \ldots, H_h$ be a list of the components of $G[S_{\mathcal C}]$ and let $H_i$ be such a component. By the above remarks, there exist at least two $C \in \mathcal C$ so that $V_{C,m} = H_i$ and $(C,m) \in P$.

% So let $K_{i1}, \ldots, K_{ik}$, $k\geq 2$, be a list of the nontrivial graphs $K_{ij} = ([|H_i|], \epsilon^G[C_j,m,\cdot,\cdot])$ canonizing $H_i$.
Notice that by the definitions of $P$ and $\prec_{M_n}$, we have $m < n$ and therefore all $\epsilon^G[C,m,\cdot,\cdot]$ with $C \in \mathcal C$ have already been defined. Let $\sim$ be the equivalence relation on $P \cap (\mathcal C \times \{m\})$ defined by $(C,m) \sim (C',m) :\Leftrightarrow V_{C,m} = V_{C',m}$. Using Lemma \ref{lem:lexDisjUnionDefable}, we obtain the lexicographic disjoint union $\omega_{\mathcal C}(x,y)$ of the lexicographic leaders of $\sim$'s equivalence classes.

%By REF HERE, the edge relation of the lexicographic leader among $\{K_{ij}\}_{j\in [k]}$ is definable by an \FPC-formula as $\lambda[H_i,\cdot,\cdot]$. Similarly, we can order $\{\lambda[H_i]\}_{i\in [h]}$ lexicographically and then define $\zeta[S_{\mathcal C}]$ as their disjoint union on consecutive intervals of $\N$.

Finally, let $<^{G_{M,n}}_M$ be the strict partial order on $V_{M,n} \modout \sim^{G_{M,n}}_M$ defined above. Let $c_1, \ldots c_k$ be the list of non-singular equivalence classes of $\sim^{G_{M,n}}_M$. Each $c_i$ is associated with a unique maximal set of incomparables $\mathcal C_i$, and $c_i = S_{\mathcal C_i}$ as sets. We aim at canonizing $G_{M,n}$ using $<^{G_{M,n}}_M$, inserting the graph defined by $\omega_{\mathcal C_i}(x,y)$ in place of each $c_i$. Here is how: each $[v] \in V_{M,n} \modout \sim^{G_{M,n}}_M$ is represented by the interval $[a+1, a + |[v]|]$, where $a$ is the number of vertices in equivalence classes strictly $<^{G_{M,n}}_M$-smaller than $[v]$. Since all vertices in $[v]$ have the same neighbors in all of $G\setminus [v]$, it is clear how to define the edge relation between $[v]$ and $G\setminus [v]$. If $[v]$ is not a singleton set, then $c_i = [v]$ for some $i$ and the edge relation on $c_i$ is given by $\omega_{\mathcal C_i}(x,y)$.

%Here is how: for each $[v] \in V_{M,n} \modout \sim^{G_{M,n}}_M$, let $p([v]) = \sum_{[u] <^{G_{M,n}}_M [v]} |[u]|$. For $x \in [|V_{M,n}|]$, let $p_x$ be the largest element in the image of $p$ such that $p_x\leq x$. Define $p^{\inv}(x) := p^{-1}(p_x)$.

%Now define $\epsilon(M,n,x,y)$ for $x\neq y$ as follows. First, assume that $p^{\inv}(x) \not\in \{c_i\}_{i\in [k]}$. Then $\epsilon(M,n,x,y)$ holds if and only if $p^{\inv}(y) \subseteq N(p^{\inv}(x))$. The definition is symmetrical if $p^{\inv}(y) \not\in \{c_i\}_{i\in [k]}$. The remaining case is when $p^{\inv}(x) = c_i$ and $p^{\inv}(y) = c_j$ for some $i,j \in [k]$. If $i\neq j$, there is no edge between $x$ and $y$ since $S_{\mathcal C_i}$ and $S_{\mathcal C_j}$ are not connected. If $i = j$, then notice that $p_x = p_y$ and we let $\epsilon(M,n,x,y) \leftrightarrow \omega_{\mathcal C_i}(x-p_x,y-p_x)$. This completes the definition of $\epsilon(M,n,x,y)$.

It is clear from the construction that $\left( [|V_{m,n}|], \epsilon^G[M,n,\cdot,\cdot] \right) \isom G[V_{M,n}]$. Also, $\epsilon(M,n,x,y)$ can be defined in \FPC for all $(M,n) \in P$ using a fixed point-operator iterating $n$ from $1$ to $|V|$. Finally, let $\varepsilon(x,y)$ be the lexicographic disjoint union of the lexicographic leaders canonizing the components of $G$, each of which is defined by some $(M,|V|) \in P$. Then $\left( [|V|], \varepsilon^G[\cdot, \cdot] \right) \isom G$, which concludes the canonization of $G$.\qed

%define a strict partial order $x <_{M,n} y$ on $V_{M,n}$ as follows. Call a max clique $A$ of $G[V_{M,n}]$ \emph{enqueued} if there is no other max clique $B$ such that $A$ and $B$ are incomparable with respect to $\prec_M$. If there are $\prec_M$-least max cliques $A,B$ so that $x\in A$, $y\in B$, then $x <_{M,n} y$ holds if and only if either
%\begin{itemize}
% \item $A \prec_M B$ or
% \item $A=B$ is enqueued and $\spn(x) < \spn(y)$.
%\end{itemize}
%Notice that if $A,B$ are not incomparable with respect to $\prec_M$, then $x\neq y$ are incomparable with respect to $<_{M,n}$ if and only if they are twins, i.e., they are represented by two copies of the same interval. Let $Z$ be the set of all $v\in V_{M,n}$ contained in some enqueued max clique. Define the edge relation $\epsilon[M,n,x,y]$ as follows:
%\[
% xy \in \epsilon[M,n,\cdot,\cdot] :\Leftrightarrow
%\]

\proof[Proof of Corollary \ref{cor:IntGraphsDefinable}] 
We claim that for the recognition of interval graphs, it is enough to check that (a) every edge of the graph $G$ is contained in some max clique which is defined by the joint neighborhood of some pair of vertices and (b) the canonization procedure as described in Section \ref{sec:CapOnIntGraphs} succeeds to produce a graph of the same size on the number sort. 

Certainly, any interval graph satisfies both conditions by the results in this paper. Conversely, assume that $G = (V,E)$ satisfies these conditions, and let $H = ([|V|],\epsilon)$ be the ordered graph defined by the canonization procedure. We can choose a bijection $\phi: V \rightarrow [|V|]$ by breaking all ties during the definition of $H$ arbritrarily. We claim that $\phi$ is an isomorphism between $G$ and $H$. Let $u,v \in V$ and suppose $uv \not\in E$. Since the sets of max cliques respectively containing $u$ and $v$ are disjoint, at no point during the canonization procedure there is an edge defined between numbers corresponding to $u$ and $v$, and hence $\phi(u)\phi(v) \not\in \epsilon$. If $uv \in E$, however, then $u,v$ are both contained in some definable max clique $C$ which is forced to appear in the relations $\prec_M$ defined by (\ref{eqn:indDef}). It is easy to see then that also $\phi(u)\phi(v) \in \epsilon$, and hence $\phi$ is an isomorphism. Finally, observe that any graph defined by the canonization procedure is an interval graph.

% It is clear from the proof of Theorem \ref{thm:captureIntGraphs} that alongside the canonical form, it is possible to construct a \FPC-formula $\prec$ defining a \emph{linear order} on the original graph's max cliques which corresponds to the linear order of the canonical form's max cliques. $\prec$ defines a strict weak order $<_G$ on the vertices of $G$ in the same way as defined in Section \ref{subsec:canLinOrder}.
% 
% Now given any graph $G$, there are three possibilities what can happen when we attempt its canonization as an interval graph. If the canonization procedure fails to produce a graph on the number sort, then $G$ is not an interval graph. If $\epsilon$ does define a graph on the number sort, then it must be in orderly one-to-one correspondence with $G$'s vertices ordered by $<_G$. If this is not the case, then $G$ is not an interval graph. If it is, then $\epsilon$ defines an isomorphic copy of $G$ on the number sort, and the \PTIME-property of being an interval graph can be verified by the Immerman-Vardi theorem. \qed
% 

\section{Conclusion}

We have proved that the class of interval graphs admits \FPC-definable canonization. Thus, \FPC captures \PTIME on the class of interval graphs, which was shown not to be the case for any of the two obvious superclasses of interval graphs: chordal graphs and incomparability graphs. The result also implies that the combinatorial Weisfeiler-Lehman algorithm solves the isomorphism problem for interval graphs.

As noted in Remark \ref{rem:alreadySTC}, the methods in this paper can be used to define \LOGSPACE-computable canonical forms for interval graphs (cf. \cite{koebler10interval}). The \FPC-canonization of the modular decomposition tree from Section \ref{subsec:generalIntGraphs} is then replaced by Lindell's \LOGSPACE-tree canonization algorithm~\cite{lindell92logspace}. This implies that the set of logspace computable intrinsic properties of interval graphs is recursively enumerable. However, \LOGSPACE is not captured by first-order logic with the symmetric transitive closure operator: any rooted tree is converted into an interval graph by connecting each vertex to all its descendants. Arguing as in Section \ref{sec:noCaptureCompGraphs}, such a capturing result would imply an analogous capturing result on trees, which is ruled out by the work of Etessami and Immerman~\cite{etessami95tree}.

%
%  Verbitsky has confirmed in private communication that using the methods of this paper, the WL algorithm can even be shown to find a canonical labeling of interval graphs in a logarithmic number of steps. By a result of Grohe and Verbitsky \cite{grohe06testing}, this implies the existence of a $\cclass{TC^1}$ parallel algorithm for the interval graph isomorphism problem, which is an improvement over the currently known $\cclass{NC^2}$ procedures.

Among the graph classes considered in this paper, the only class whose status is not settled with respect to \FPC-canonization is the class of chordal comparability graphs. While it appears that the methods employed for chordal incomparability graphs here do not carry over (see Remark \ref{remark:doesntWorkGenerally}), I believe I have found a different solution, which will be contained in the journal version of this paper.

So far, little is known about logics capturing complexity classes on classes of graphs which are defined by a (finite or infinite) list of forbidden induced subgraphs. This paper makes a contribution in this direction. It seems that chordal graphs, even though they do not admit \FPC-canonization themselves, can often be handled effectively in fixed-point logic as soon as additional properties are satisfied (being a line graph, incomparability or comparability graph). It would be instructive to unify these properties. In this context, I would also like to point to Grohe's conjecture~\cite{grohe09fixed-point} that \FPC captures \PTIME on the class of claw-free chordal graphs.

%Finally, methods need to be developed in order to deal with non-chordal graph classes. It seems possible that the capturing result for interval graphs could be extended to a capturing result for the superclass of circular-arc graphs, which would be a first step in that direction. A closer investigation of graphs of bounded rank-width~\cite{oum05rank-width} seems another natural inroad for finding out more about such graph classes.

\section*{Acknowledgement}
I would like to thank Martin Grohe for bringing the question of capturing \PTIME on interval graphs to my attention and for helpful discussions on the subject.

%% in general the use of bibtex is encouraged

%\nocite{ex1,ex2}
\bibliographystyle{plain}
\bibliography{references}

\begin{thebibliography}{10}

\bibitem{booth76testing}
Kellogg~S. Booth and George~S. Lueker.
\newblock Testing for the consecutive ones property, interval graphs, and graph
  planarity using pq-tree algorithms.
\newblock {\em J. Comput. Syst. Sci.}, 13(3):335--379, 1976.

\bibitem{brandstaedt99graph}
Andreas Brandst\"adt, Van~Bang Le, and Jeremy~P. Spinrad.
\newblock {\em Graph Classes: a Survey}.
\newblock Monographs on Discrete Mathematics and Applications. SIAM, 1999.

\bibitem{cai92optimal}
J.~Cai, M.~F{\"u}rer, and N.~Immerman.
\newblock An optimal lower bound on the number of variables for graph
  identification.
\newblock {\em Combinatorica}, 12(4):389--410, 1992.

\bibitem{dawar07power}
Anuj Dawar and David Richerby.
\newblock The power of counting logics on restricted classes of finite
  structures.
\newblock In Jacques Duparc and Thomas~A. Henzinger, editors, {\em CSL}, volume
  4646 of {\em Lecture Notes in Computer Science}, pages 84--98. Springer,
  2007.

\bibitem{diestel06graphtheory}
Reinhard Diestel.
\newblock {\em Graph Theory}, volume 173 of {\em Graduate Texts in
  Mathematics}.
\newblock Springer, 3rd edition, 2006.

\bibitem{ebbinghaus99finite}
H.-D. Ebbinghaus and J.~Flum.
\newblock {\em Finite model theory}.
\newblock Springer-Verlag, 2nd edition, 1999.

\bibitem{ebbinghaus94mathematical}
H.-D. Ebbinghaus, J.~Flum, and W.~Thomas.
\newblock {\em Mathematical Logic}.
\newblock Springer-Verlag, 2nd edition, 1994.

\bibitem{etessami95tree}
Kousha Etessami and Neil Immerman.
\newblock Tree canonization and transitive closure.
\newblock In {\em LICS '95}, pages 331--341. IEEE Computer Society, 1995.

\bibitem{gilmore64characterization}
P.~C. Gilmore and A.~J. Hoffman.
\newblock A characterization of comparability graphs and of interval graphs.
\newblock {\em Canad. J. Math.}, 16:539--548, 1964.

\bibitem{golumbic04algorithmic}
M.C. Golumbic.
\newblock {\em Algorithmic Graph Theory and Perfect Graphs (Annals of Discrete
  Mathematics, Vol 57)}.
\newblock North-Holland Publishing Co., Amsterdam, The Netherlands, 2004.

\bibitem{graedel07finite}
E.~Gr\"adel, P.G. Kolaitis, L.~Libkin, M.~Marx, J.~Spencer, M.Y. Vardi,
  Y.~Venema, and S.~Weinstein.
\newblock {\em Finite Model Theory and Its Applications}.
\newblock Texts in Theoretical Computer Science. Springer, Berlin, Germany,
  2007.

\bibitem{griggs79extremal}
J.R. Griggs and D.B. West.
\newblock Extremal values of the interval number of a graph.
\newblock {\em SIAM J. Alg. Discrete Methods}, 1:1--7, 1979.

\bibitem{gro98a}
M.~Grohe.
\newblock Fixed-point logics on planar graphs.
\newblock In {\em LICS '98}, pages 6--15. IEEE Computer Society, 1998.

\bibitem{grohe08definable}
M.~Grohe.
\newblock Definable tree decompositions.
\newblock In {\em LICS '08}, pages 406--417. IEEE Computer Society, 2008.

\bibitem{grohe09fixed-point}
M.~Grohe.
\newblock Fixed-point definability and polynomial time on chordal graphs and
  line graphs.
\newblock {\em arXiv}, 1001.2572, 2010.

\bibitem{gromar99}
M.~Grohe and J.~Mari{\~{n}}o.
\newblock Definability and descriptive complexity on databases of bounded
  tree-width.
\newblock In {\em ICDT '99}, volume 1540 of {\em LNCS}, pages 70--82. Springer,
  1999.

\bibitem{grohe10fixed}
Martin Grohe.
\newblock Fixed-point definability and polynomial time on graphs with excluded
  minors.
\newblock In {\em LICS '10 (this conference)}, 2010.

\bibitem{gurevich85fixed-point}
Yuri Gurevich and Saharon Shelah.
\newblock Fixed-point extensions of first-order logic.
\newblock In {\em 26th Annual Symposium on Foundations of Computer Science},
  pages 346--353, 1985.

\bibitem{habib00lex-bfs}
Michel Habib, Ross~M. McConnell, Christophe Paul, and Laurent Viennot.
\newblock Lex-{BFS} and partition refinement.
\newblock {\em Theor. Comput. Sci.}, 234(1-2):59--84, 2000.

\bibitem{hsu99fast}
Wen-Lian Hsu and Tze-Heng Ma.
\newblock Fast and simple algorithms for recognizing chordal comparability
  graphs and interval graphs.
\newblock {\em SIAM J. Comput.}, 28(3):1004--1020, 1999.

\bibitem{immerman82bounds}
N.~Immerman.
\newblock Upper and lower bounds for first-order expressibility.
\newblock {\em Journal of Computer and System Sciences}, 25:76--98, 1982.

\bibitem{immerman99descriptive}
N.~Immerman.
\newblock {\em Descriptive Complexity}.
\newblock Springer, 1999.

\bibitem{koebler10interval}
Johannes K\"obler, Sebastian Kuhnert, Bastian Laubner, and Oleg Verbitsky.
\newblock Interval graphs: Canonical representation in logspace.
\newblock In {\em ICALP}, in print, 2010.

\bibitem{kreutzer04equivalence}
Stephan Kreutzer.
\newblock Expressive equivalence of least and inflationary fixed-point logic.
\newblock {\em Annals of Pure and Applied Logic}, 130(1-3):61--78, 2004.

\bibitem{lekkerkerker62representation}
C.~G. Lekkerkerker and J.~Ch. Boland.
\newblock Representation of a finite graph by a set of intervals on the real
  line.
\newblock {\em Fundamenta Mathematicae}, 51:45--64, 1962.

\bibitem{lindell92logspace}
S.~Lindell.
\newblock A logspace algorithm for tree canonization.
\newblock In {\em STOC}, pages 400--404. ACM, 1992.

\bibitem{lueker79linear}
George~S. Lueker and Kellogg~S. Booth.
\newblock A linear time algorithm for deciding interval graph isomorphism.
\newblock {\em J. ACM}, 26(2):183--195, 1979.

\bibitem{moehring84algorithmic}
R.H. M\"ohring.
\newblock {\em Graphs and Order}, volume 147 of {\em NATO ASI Series C,
  Mathematical and Physical Sciences}, pages 41--102.
\newblock D. Reidel, 1984.

\bibitem{Uehara08simple}
Ryuhei Uehara.
\newblock Simple geometrical intersection graphs.
\newblock In {\em WALCOM}, pages 25--33, 2008.

\bibitem{vardi82complexity}
M.Y. Vardi.
\newblock The complexity of relational query languages.
\newblock In {\em STOC '82}, pages 137--146, 1982.

\bibitem{zhang94algorithm}
Peisen Zhang, Eric~A. Schon, Stuart~G. Fischer, Eftihia Cayanis, Janie Weiss,
  Susan Kistler, and Philip~E. Bourne.
\newblock An algorithm based on graph theory for the assembly of contigs in
  physical mapping of {DNA}.
\newblock {\em Bioinformatics}, 10(3):309--317, 1994.

\end{thebibliography}

\end{document}